\documentclass[12pt, draftclsnofoot, onecolumn]{IEEEtran}
\usepackage{graphicx,amssymb,amsmath}
\usepackage{multicol}
\usepackage[noadjust]{cite}
\usepackage{setspace}
\usepackage{stfloats}
\usepackage{cite}
\usepackage{amsthm}
\usepackage{flushend}
\usepackage{cases,subeqnarray}
\usepackage{bm,multirow,bigstrut}
\usepackage{textcomp}
\usepackage{latexsym,bm}
\usepackage{xcolor}
\usepackage{mathtools}
\usepackage{dsfont}
\usepackage{extarrows}
\usepackage{algorithm}
\usepackage{algpseudocode}
\usepackage{amsmath}
\usepackage{amsthm}
\usepackage{booktabs}

\theoremstyle{plain}

\theoremstyle{plain}

\newtheorem{myDef}{Definition}

\IEEEoverridecommandlockouts

\begin{document}

\title{Federated Learning-Based Cell-Free Massive MIMO System for Privacy-Preserving}
\author{Jiayi~Zhang,~\IEEEmembership{Senior Member,~IEEE}, Jing Zhang, Derrick~Wing~Kwan Ng,~\IEEEmembership{Fellow,~IEEE,}
and Bo Ai,~\IEEEmembership{Fellow,~IEEE}
\thanks{J. Zhang and J. Zhang are with the School of Electronic and Information Engineering, Beijing Jiaotong University, Beijing, China.  (email: jiayizhang@bjtu.edu.cn).}
\thanks{D.~W.~K. Ng is with School of Electrical Engineering and Telecommunications, University of New South Wales, Sydney, N.S.W., Australia (email: w.k.ng@unsw.edu.au).}
\thanks{B. Ai is with State Key Laboratory of Rail Traffic Control and Safety, Beijing Jiaotong University, Beijing 100044, China (email: boai@bjtu.edu.cn).}
}
\maketitle

\begin{abstract}
Cell-free massive MIMO (CF mMIMO) is a promising next generation wireless architecture to realize federated learning (FL). However, sensitive information of user equipments (UEs) may be exposed to the involved access points or the central processing unit in practice. To guarantee data privacy, effective privacy-preserving mechanisms are defined in this paper. In particular, we demonstrate and characterize the possibility in exploiting the inherent quantization error, caused by low-resolution analog-to-digital converters (ADCs) and digital-to-analog converters (DACs), for privacy-preserving in a FL CF mMIMO system.
Furthermore, to reduce the required uplink training time in such a system, a stochastic non-convex design problem \textcolor{black}{that jointly optimizing the transmit power and the data rate is formulated.}
To address the problem at hand, we propose a novel power control method by utilizing the successive convex approximation approach to obtain a suboptimal solution.
Besides, an asynchronous protocol is established for mitigating the straggler effect to facilitate FL.
Numerical results show that compared with the conventional full power transmission, adopting the proposed power control method can effectively reduce the uplink training time under various practical system settings.
Also, our results unveil that our proposed asynchronous approach can reduce the waiting time at the central processing unit for receiving all user information, as there are no stragglers that requires a long time to report their local updates.
\end{abstract}

\begin{IEEEkeywords}
Cell-free massive MIMO, federated learning, power control, differential privacy.
\end{IEEEkeywords}


\IEEEpeerreviewmaketitle

\section{Introduction}
Massive multiple-input multiple-output (MIMO) is an unprecedented technique to increase both the spectral efficiency (SE) and energy efficiency (EE) of communication systems.
As a result, massive MIMO has already been utilized in practical cellular systems \cite{[1],[3],[4],[5],9205230}.
However, some serious concerns have recently been raised about data privacy.
Indeed, mobile devices nowadays are often equipped with high computing capabilities enabling them to collect and process large amounts of data \cite{[11],[12]}.
Specifically, numerous applications perform data preprocessing and classification for predicting possible future events via using various machine learning technologies \cite{ML1,ML2}.
The vast amount of data of devices is generally collected for numerous private applications carrying sensation information and thus naturally causes privacy concerns.
On the other hand, it is generally challenging to transmit all the data to a central processing unit (CPU) for training a deep learning model. Besides, due to the limited resources of wireless systems, sending a large amount of data through wireless link is not always possible as it would introduce expensive communication costs and exceedingly long communication delays.

\subsection{Related Works}
In order to address the above challenges, it is necessary to design a novel machine learning (ML) technology such that each user equipment (UE) can be trained locally based on the data it collects and collaboratively establish a shared global learning model.
One of the most promising decentralized learning methods to achieve this goal is FL \cite{wahab2021federated,9718315}.
In particular, multiple UEs are allowed to jointly train a global ML model without having to exchange raw data among them or transfer their data to the CPU \cite{amiri2020federated}.
Specifically, the CPU first broadcasts the latest global model to all the participating UEs.
Next, the UEs calculate the corresponding local update based on the available data and then send their local models back to the CPU.
Repeat these steps until a certain level of global model accuracy is reached.
In this way, only local model parameters are exchanged, thereby reducing the required communication signaling overhead.

Despite raw data sharing wireless channel can be avoided via FL, UEs' sensitive information can still be possibly revealed through any form of the leaked information.
For example, a malicious CPU can perform a model inversion attack \cite{fredrikson2015model} to infer the presence of individual data samples.
Moreover, other adversaries can apply differential attacks to the wireless communication phase that performs data exchanged between the CPU and distributed access points (APs) \cite{wahab2021federated}. In the literature, there are three popular techniques for maintaining privacy, including anonymization, data encryption and differential privacy (DP) \cite{sweeney2002k,wang2018privacy,dwork2011firm}, with different drawback. For instance, anonymization strategies do not guarantee complete level of protection from adversaries; cryptographic techniques are computationally expensive. In contrast, differential privacy is easy to implement and provides provable privacy guarantee.
Specifically, DP prevents the sensitive information of UEs from being easily exposed even if the CPU/adversaries can access the model parameters and acquire the knowledge of the adopted training mechanism \cite{ma2020safeguarding}.
In fact, one appealing approach to realize DP is via dedicated noise injection \cite{liu2020privacy}.
The main idea of this approach is to deliberately introduce some noises to the uploaded local model updates such that the CPU/adversaries cannot infer any sensitive information from exploiting the actual data.
Recently, remarkable efforts have been made to investigate DP mechanisms for wireless FL through artificial noise injection.
For instance, in \cite{wei2020federated}, Gaussian noise was added to the local updated data and the power control was applied to realize different levels of DP protection.
Besides, the results in \cite{liu2020privacy} and \cite{koda2020differentially} showed that the inherent channel noise can be exploited for guaranteeing DP FL.
Indeed, by deploying a proper power control, one can harness the channel noise to achieve privacy for free.
Also, in \cite{aghdam2021privacy}, the inherent hardware-induced distortion was exploited to facilitate local model updates and a power allocation strategy was proposed to provide guaranteed DP.
On the other hand, the existence of straggler effect also creates a bottleneck in realizing effective FL in wireless networks \cite{vu2020user}.
By definition, the CPU needs to wait until it receives training updates from all UEs before processing any next steps.
Therefore, some straggler UEs with unfavorable links may greatly slow down the entire FL process and reduce its practicality \cite{vu2020user,wu2020safa}.

\subsection{Contributions}
All the aforementioned works, e.g. \cite{liu2020privacy,wei2020federated,koda2020differentially}, only assume simple wireless environments, e.g. Gaussian channels with additive white noise, which does not consider the fluctuation of practical wireless channels.
Also, in practice, when the number of UEs increases, the required training time could be significantly prolonged.
In such scenarios, to serve a large number of UEs via the same time/frequency resources, cell-free massive multiple-input multiple-output (CF mMIMO) systems have been applied for supporting FL \cite{zhang2020prospective}.
Herein, multiple distributed APs are connected to the CPU through capacity-unlimited fronthaul links to serve UEs coherently.
Since the large number of APs can provide a rich macro-diversity gain for ensuring uniform received power strength, the performance of CF mMIMO supported FL is less prone to UEs with weak communication links.
However, the current literature focuses on cell-free massive MIMO systems implemented with federate learning is still limited.
A scheme for CF mMIMO networks to support any FL framework was proposed in \cite{vu2020cell} for the first time. An optimization problem was also formulated to jointly optimize the local accuracy, transmit power, data rate, and users¡¯ processing frequency, and is solved by employing the online successive convex approximation approach.
Also, a UE selection approach was proposed in \cite{vu2021straggler} to mitigate the straggler effect with UE sampling for FL in CF mMIMO networks. It selects only a small subset of UEs for participating in one FL process.
In \cite{vu2021does}, a novel scheme that asynchronously executes the iterations of FL processes was designed for multicasting downlink and conventional uplink transmission protocols.
However, the privacy-preserving is not considered in \cite{vu2020cell,vu2021straggler,vu2021does}.
In \cite{xu2021privacy}, the authors developed and analyzed a privacy-preserving channel estimation schemes in CF mMIMO systems.
Yet, the design in \cite{xu2021privacy} only provides the data privacy protection during the channel estimation phase, therefore the data information transmitted by the UE in the payload data transmission phase still with high potential of leakage.
Besides, practical digital CF mMIMO communication systems adopt low-resolution analog-to-digital converters (ADCs) and digital-to-analog converters (DACs) to reduce associated power consumption and hardware cost \cite{hu2019cell}. In fact, these inherent noise can be exploited to enhance DP that is somewhat overlooked in the literature.

Motivated by the above discussion, we consider a practical CF mMIMO supported FL framework and demonstrate that the inherent quantization noise caused by low-resolution ADCs and DACs in CF mMIMO can be exploited as a useful privacy-preserving mechanism. Our contributions are listed as follows:
\begin{itemize}
  \item First, we capitalize the quantization noise introduced by the low-resolution ADCs and DACs to prevent the CPU/adversaries from exploiting the actual local gradient updates to infer sensitive information and hence, realize privacy-preserving. Within this proposed framework, we derive an upper-bound to characterize the privacy violation probability and adopt it for formulating a privacy preservation condition.
  \item \textcolor{black}{Then, we provide the closed-from convergence analysis of the DP mechanism, taking into account the quantization noise, the length of the local updates, and the total data size. It can be observed from the upper bound of the average optimality gap that the noise added in the initial iterations is less damaging to the final optimality gap than that added in later iterations. Besides, the initial optimality gap decays geometrically as the number of iteration increases.
      Note that the quantization noise is exploited for privacy protection, therefore, different quantization accuracies can realize different DP protection levels.}
  \item \textcolor{black}{Third, in order to minimize the uplink training time of CF mMIMO-supported-FL, we formulate a stochastic nonconvex optimization problem that jointly optimizes the transmit power and the data rate, subject to the practical constraints on energy consumption of the UEs with different quantization accuracies of the ADCs and DACs. Our numerical results reveal that the proposed power control method can effectively reduce the uplink training time under different privacy protection levels. Besides, our power control method still performs well over various baseline schemes for different number of APs and UEs.}
  \item Finally, we propose an asynchronous FL protocol to alleviate the straggler effect with two simple parameters, \textit{lag tolerance} and \textit{lag percent}, respectively. We also compare the performance of synchronous FL and asynchronous FL and the empirically analyze the impacts caused by lag tolerance and lag percent.
\end{itemize}

\emph{Notations:} Throughout the paper, $\mathbb{R}$ and $\mathbb{C}$ represents the sets of all real and complex values, respectively. We denote a complex zero-mean normal distribution with variance $\sigma ^2$ by ${{\cal N}_{\mathbb{C}}}\left( {0,{\sigma ^2}} \right)$. The cardinality of a set $\mathcal{A}$ is denoted by $\left| {\cal A} \right|$. Furthermore, the difference between two sets is defined as ${\cal A}' - {\cal A}'' = \left\{ {\left. x \right|x \in {\cal A}',x \notin {\cal A}''} \right\}$. The union of sets ${{\cal A}_1}, \cdots ,{{\cal A}_K}$ is represented by ${\cal A} = \bigcup\nolimits_{k = 1}^K {{{\cal A}_k}}$.
Boldfaced lower-case letters, e.g., $\bf{a}$, represents the vectors, ${\bf{a}}^H$, ${\bf{a}}^*$, and $\left\| {\bf{a}} \right\|$ denote Hermitian transpose, conjugate, and Euclidean norm of $\bf{a}$, respectively.

\section{System Model}\label{system_model}
\begin{figure}[t!]
\centering
\includegraphics[scale = 0.8]{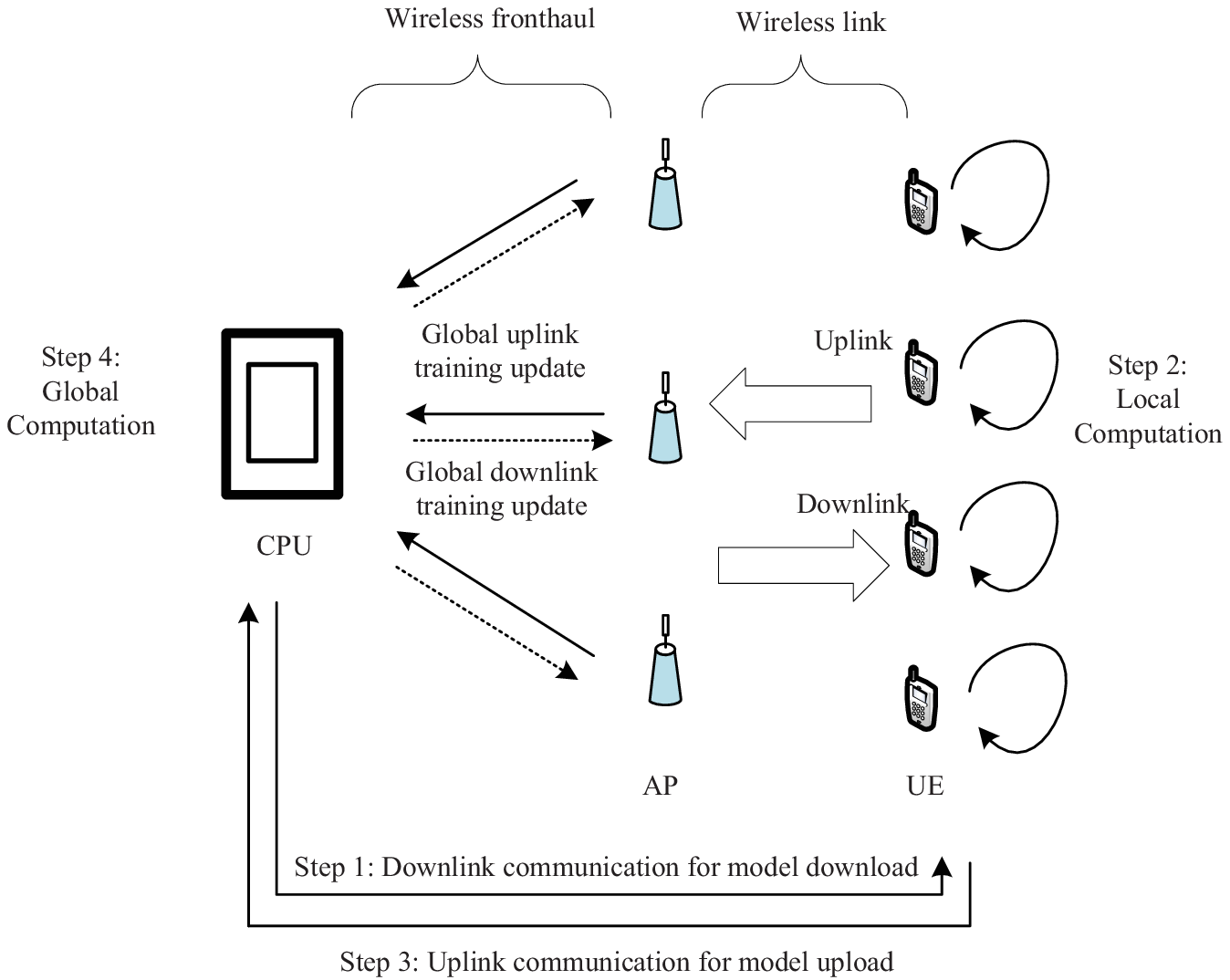}
\textcolor{black}{\caption{Illustrations of a CF mMIMO system supported FL with low-resolutions ADCs equipped at the APs.}}
\label{fig_systemmodel}
\end{figure}

As shown in Fig. \ref{fig_systemmodel},
we consider a CF mMIMO supported FL system consisting of $K$ single-antenna UEs and $L$ single-antenna APs \cite{ngo2017cell}.
All APs and UEs are randomly located in an $D \times D$ area.
Each UE is served by all the APs over the same time/frequency resources.
A CPU is connected to all APs via ideal wireless fronthaul. UE $k$, $k \in \left\{ {1, \cdots ,K} \right\}$, is equipped with its own local dataset ${{\cal B}_k} \buildrel \Delta \over = \left\{ {\left( {{{\bf{x}}_{kn}},{y_{kn}}} \right)} \right\}_{n = 1}^{{B_k}}$, where $B_k$ is the data size and ${\left( {{{\bf{x}}_{kn}},{y_{kn}}} \right)}$ is the corresponding $n$th data sample. The objective of FL is to find an $d \times 1$ optimal model vector ${\bf{w}}$ that minimizes the global loss function \cite{aghdam2021privacy}:
\begin{equation}\label{(1)}
\mathop {\text{minimize} }\limits_{\bf{w}} {\rm{    }}F\left( {\bf{w}} \right) = \frac{1}{K}\sum\limits_{i = 1}^K {{F_i}} \left( {\bf{w}} \right),
\end{equation}
where ${B_{{\rm{tot}}}} = \sum\limits_{i = 1}^K {{B_k}}$ and ${{F_k}\left( {\bf{w}} \right)}$ is the local loss function which is given by
\begin{equation}\label{(2)}
{F_k}\left( {\mathbf{w}} \right) = \frac{1}{{{B_k}}}\sum\nolimits_{\left( {{{\mathbf{x}}_k},{y_k}} \right) \in {\mathcal{B}_k}} {f\left( {{\mathbf{w}},{{\mathbf{x}}_k},{y_k}} \right)},
\end{equation}
where ${f\left( {{\bf{w}};{{\bf{x}}_k},{y_k}} \right)}$ is the sample-wise loss function that quantifies the prediction error of the model ${\bf{w}}$ on the training samples ${{\bf{x}}_{kn}}$ with respect to the labels $y_{kn}$.

\subsection{Learning Protocol}
In order to address problem (\ref{(1)}), we apply the distributed stochastic gradient descent (SGD) \cite{SGD1,SGD2} at the CPU and the UEs. Note that the CPU and the UEs, respectively, act as the central server and the clients in the general FL framework. The APs with CF mMIMO are only used to relay the training updates between the CPU and the UEs. The specific procedure is summarized as follows.
\begin{itemize}
  \item {\bf{Step 1: Downlink communication for model download}}. The CPU broadcasts the current model, i.e., ${\bf{w}}^t$, to all the UEs, where $t$ represents the communication round, $t = 1, \cdots ,T$.
  \item {\bf{Step 2: Local computation.}} Each UE computes the gradient of the local loss function in (\ref{(2)}) via
        \begin{equation}\label{(3)}
        \nabla {F_k}\left( {{{\bf{w}}^t}} \right) = \frac{1}{{{B_k}}}\sum\limits_{\left( {{{\bf{x}}_k},{y_k}} \right) \in {{\cal B}_k}} \nabla {f\left( {{{\bf{w}}^t},{{\bf{x}}_k},{y_k}} \right)}, \forall t,
        \end{equation}
        where $\nabla {F_k}\left( {{{\bf{w}}^t}} \right)$ is the gradient of ${F_k}\left( {{{\bf{w}}^t}} \right)$.
  \item {\bf{Step 3: Uplink communication for model upload}}. The UEs send the gradient of the local loss function to the CPU utilizing the same time and frequency resources.
  \item {\bf{Step 4: Global computation}}. Based on the received signal, the CPU obtains an estimated $\widehat {\nabla F}\left( {{{\bf{w}}^t}} \right)$ of the global gradient by computing
        \begin{equation}\label{(4)}
        \nabla F\left( {{{\bf{w}}^t}} \right) = \frac{1}{{{B_{{\rm{tot}}}}}}\sum\limits_{i = 1}^K {\nabla {F_i}\left( {{{\bf{w}}^t}} \right)}.
        \end{equation}
        Then, the CPU updates the current global model as
        \begin{equation}
        {{\bf{w}}^{t + 1}} = {{\bf{w}}^t} - \eta \widehat {\nabla F}\left( {{{\bf{w}}^t}} \right),
        \end{equation}
        where ${\eta}$ denotes the learning rate.
\end{itemize}
Note that Steps 2 to 4 are repeated until a convergence criterion is met.

\subsection{Communication Model}
The channel coefficient between AP $l$ and UE $k$ is denoted as $h_{kl} \in \mathbb{C}$, which is modeled as ${h_{kl}} = \sqrt {{\beta _{kl}}} {g_{kl}}$, where $\beta \in \mathbb{R}$ represents the large-scale fading and $g_{kl} \in \mathbb{C}$ represents the small-scale fading coefficient, respectively \cite{9529197}. We adopt the block fading model where $h_{kl}$ is a constant in each time-frequency block. Without loss of generality, \textcolor{black}{each} block contains ${\tau _c}$ channel uses, which consists of $\tau_p$ channel uses dedicated for acquiring the channel state information (CSI) and $\tau_c-\tau_p$ channel uses for the uncoded transmission of the $d$-dimensional gradient vector \cite{liu2020privacy}. Besides, we assume that perfect CSI is available at the APs \cite{liu2020privacy}. In the following, we derive the uplink training expressions when consider inherent noise induced by the low-resolution ADCs at the APs or low-resolution DACs at the UEs.
\subsubsection{Low-resolution ADCs equipped at the APs}
At communication round $t$ of Step 4, the received signal ${\mathbf{y}}_l^t$ at AP $l$ is given as
\begin{equation}
{\mathbf{y}}_l^t = \sum\limits_{i = 1}^K {\sqrt {\frac{{p_i^t}}{{{{\left\| {{\mathbf{s}}_i^t} \right\|}^2}}}} {h_{il}}{\mathbf{s}}_i^t}  + {{\mathbf{n}}_l} ,
\end{equation}
where
\begin{equation}\label{s}
{\mathbf{s}}_i^t = \left| {{{{\mathcal{B}}}_i}} \right|\nabla {F_i}\left( {{{\mathbf{w}}^t}} \right),\;\;\;\forall i \in \left\{ {1, \cdots ,K} \right\},
\end{equation}
where ${p_k^t}$ denotes the transmit power for UE $i$, ${{\mathbf{n}}_l}$ is the additive noise with independent ${\mathcal{N}_\mathbb{C}}\left( {0,{\sigma ^2}{{\mathbf{I}}_d}} \right)$, and ${\sigma ^2}$ is the noise power per antenna.
We adopt the linear additive quantization noise model \cite{fletcher2007robust} to capture the quantization loss and the noise caused by low-resolution ADCs, which yields
\begin{equation}
\mathcal{Q}\left( {{\mathbf{y}}_l^t} \right) = \alpha \sum\limits_{i = 1}^K {\sqrt {\frac{{p_i^t}}{{{{\left\| {{\mathbf{s}}_i^t} \right\|}^2}}}} {h_{il}}{\mathbf{s}}_i^t}  + \alpha {{\mathbf{n}}_l} + {\mathbf{n}}_l^{{\text{uq}}},
\end{equation}
where $\alpha  = \frac{{\pi \sqrt 3 }}{2}{2^{ - 2b}}$ is a linear gain depending on the number of quantization bits adopted in ADCs, $b$, and ${\mathbf{n}}_l^{{\text{uq}}}$ represents the additive Gaussian noise with covariance matrix\footnote{The linear gain $\alpha$ for different quantization bits can be approximated according to \cite{hu2019cell}. In general, one can adopt $b = 10$ to mimic the perfect ADCs case.} \cite{zhang2017performance}
\begin{equation}
{{\mathbf{R}}_{{\mathbf{n}}_l^{{\text{uq}}}}^t} = \alpha \left( {1 - \alpha } \right)\left( {\sum\limits_{i = 1}^K {{p_i^t}{\beta _{il}}}  + {\sigma ^2}} \right){{\mathbf{I}}_d}.
\end{equation}

We consider a fully distributed CF mMIMO system, in which the data detection is performed at the APs \cite{9622183,zheng2022cell}. When applying the maximum-ratio combining (MRC) for low computational complexity, the local processed signal for UE $k$ at AP $l$ at communication round $t$ is given as
\begin{equation}
{\mathbf{\hat s}}_{kl}^t = \alpha \sum\limits_{i = 1}^K {\sqrt {\frac{{p_i^t}}{{{{\left\| {{\mathbf{s}}_i^t} \right\|}^2}}}} h_{kl}^*{h_{il}}{\mathbf{s}}_i^t}  + \alpha h_{kl}^*{{\mathbf{n}}_l} + h_{kl}^*{\mathbf{n}}_l^{{\text{uq}}}.
\end{equation}
Then, the APs convey the local processed signal to the CPU. The received signal from all the APs at the CPU is given as
\begin{equation}\label{(10)}
{\mathbf{r}}_k^t  = \sum\limits_{l = 1}^L {{\mathbf{\hat s}}_{kl}^t} = \alpha \sum\limits_{l = 1}^L {\sum\limits_{i = 1}^K {\sqrt {\frac{{p_i^t}}{{{{\left\| {{\mathbf{s}}_i^t} \right\|}^2}}}} h_{kl}^*{h_{il}}{\mathbf{s}}_i^t} }  + {\bm{\omega }}_k^t,
\end{equation}
where ${\bm{\omega }}_k^t = \alpha \sum\limits_{l = 1}^L {h_{kl}^*{{\mathbf{n}}_l}}  + \sum\limits_{l = 1}^L {h_{kl}^*{\mathbf{n}}_l^{{\text{uq}}}}$ is the effective noise distributed according to ${\mathcal{N}_\mathbb{C}}\left( {0,{{\left( {m_k^t} \right)}^2}{{\mathbf{I}}_d}} \right)$, with ${{\left( {m_k^t} \right)}^2} = {{\alpha ^2}\sum\limits_{l = 1}^L {{\beta _{kl}}{\sigma ^2}}  +  + \alpha \left( {1 - \alpha } \right)\sum\limits_{l = 1}^L {{\beta _{kl}}} \left( {\sum\limits_{i = 1}^K {{p_i}{\beta _{il}}}  + {\sigma ^2}} \right)}$.

Now, we rigorously analyze the performance by using the achievable rates. According to \cite{ngo2017cell,9737367}, the achievable rate of UE $k$, ${R_k}$, is
\begin{align}\label{R}
&{R_k} \leqslant {r_k},\\
&{r_k} = \left( {1 - \frac{{{\tau _p}}}{{{\tau _c}}}} \right)B{\log _2}\left( {1 + {\text{SIN}}{{\text{R}}_k}} \right),
\end{align}
where $B$ is the bandwidth and
\begin{equation}\label{R_qu}
{\text{SIN}}{{\text{R}}_k} = \frac{{{p_k}{A_k}}}{{\sum\nolimits_{i \ne k}^K {{p_i}} {B_{ki}} + {C_k} + {D_k} + \sum\nolimits_{i = 1}^K {{p_i}} {E_{ki}}}},
\end{equation}
where
\begin{align}
&{A_k} = {\alpha ^2}{\left| {\sum\limits_{l = 1}^L {{\beta _{kl}}} } \right|^2}, \;\;\;{B_{ki}} = {\alpha ^2}\sum\limits_{l = 1}^L {{\beta _{kl}}{\beta _{il}}},\notag\\
&{C_k} = {\alpha ^2}{\sigma ^2}\sum\limits_{l = 1}^L {{\beta _{kl}}},\;\;\;{D_k} = d\alpha \left( {1 - \alpha } \right)\sum\limits_{l = 1}^L {{\beta _{kl}}} {\sigma ^2},\notag\\
&{E_{ki}} = d\alpha \left( {1 - \alpha } \right)\sum\limits_{l = 1}^L {{\beta _{il}}{\beta _{kl}}}.
\end{align}
According to \cite{vu2020cell}, the uplink latency of UE $k$ in each iteration involves the transmission delay of sending the global uplink training update from it to the APs and from the APs to the CPU, i.e.,
\begin{equation}\label{ut1}
{t_k} = \frac{S}{{{R_k}}},\;\;\;
{t_l} = \frac{{KS}}{{\sum\limits_{k = 1}^K {{R_k}} }},
\end{equation}
respectively, where $S$ is the data size.
Therefore, the total uplink training time is
\begin{equation}\label{(16)}
T_{\text{time}} = \mathop {\max }\limits_k \left\{ {\frac{ST}{{{R_k}}}} \right\} + \frac{{KST}}{{\sum\limits_{k = 1}^K {{R_k}} }}.
\end{equation}

\subsubsection{Low-resolution DACs at the UEs}
With low-resolution DACs equipped at the UEs, the signal sent by UE $i$, ${{{\bf{s}}_i}}$, to the APs is given as
\begin{equation}
{\cal Q}\left( {{{\bf{s}}_i}} \right) = {\zeta}{{{\bf{s}}_i}} + {\bf{n}}_i^{\rm{q}},\forall i \in \left\{ {1, \cdots ,K} \right\},
\end{equation}
\textcolor{black}{where $\zeta  = \frac{{\pi \sqrt 3 }}{2}{2^{ - 2b}}$ is a linear gain depending on the number of quantization bits adopted in DACs, $b$,} and the elements of ${{{\bf{n}}_i}}^{\rm{q}}$ are i.i.d. ${\cal C}{\cal N}\left( {0,{\zeta}\left( {1 - {\zeta}} \right){p_i}} \right)$ random variables \cite{DAC1,DAC2}.
Besides, according to (\ref{(16)}), in synchronous FL \cite{syn1, syn2}, which refers to the CPU needs to wait for receiving the training updates from all the UEs, the straggler UEs with unfavorable links may greatly slow down the entire FL process and reduce its practicality.

On the other hand, in asynchronous FL, even if all local model updates are not received, the aggregate model can be derived.
Therefore, in order to mitigate the straggler effect, the asynchronous communication mode is considered. Then, the connection relationship between the APs and the UEs can be expressed as
\begin{equation}
{d_{il}^t} = \left\{ {\begin{array}{*{20}{c}}
{1,l \in {{\cal M}_i}},\\
{0,l \notin {{\cal M}_i}},
\end{array}} \right.\;\;\;t = 1, \cdots ,T,
\end{equation}
where
\begin{align}
{{\cal M}_i^t} = \left\{ {l:{d_{il}^t} = 1,l \in \left\{ {1, \cdots L} \right\}} \right\}.
\end{align}
Note that ${d_{il}^t = 1}$ if the $l$th AP is allowed to serve UE $i$ at communication round $t$ and 0 otherwise. Therefore, ${{\cal M}_i^t}$ denotes the subset of APs that serve UE $i$ at communication round $t$.
Therefore, at communication round $t$, the transmitted signal from UE $k$ processed by the low-precision DAC is given as
\textcolor{black}{
\begin{align}
&{\bf{\mathord{\buildrel{\lower3pt\hbox{$\scriptscriptstyle\smile$}}
\over s} }}_k^t \!\! = \!\! {\cal Q}\left( {{\bf{s}}_k^t} \right) \!\!=\!\! {\zeta}{\bf{s}}_k^t + {\left( {{\bf{n}}_k^{\rm{q}}} \right)^t} .
\end{align}}
The received signal ${\mathbf{y}}_l^t$ at AP $l$ is
\textcolor{black}{\begin{align}
{\bf{y}}_l^t &= \sum\limits_{i = 1}^K {d_{il}^t}{h_{il}^t{\bf{\mathord{\buildrel{\lower3pt\hbox{$\scriptscriptstyle\smile$}}
\over s} }}_i^t}  + {\bf{n}}_l^t = \sum\limits_{i = 1}^K {{\zeta }{d_{il}^t}h_{il}^t{\bf{s}}_i^t}  + \sum\limits_{i = 1}^K {d_{il}^t}{h_{il}^t{{\left( {{\bf{n}}_k^{\rm{q}}} \right)}^t}}  + {\bf{n}}_l^t
= \sum\limits_{i = 1}^K {{\zeta }{d_{il}^t}h_{il}^t{\bf{s}}_i^t}  + {\bm{\omega }}_k^t,
\end{align}}
where ${\bm{\omega }}_k^t = \sum\limits_{i = 1}^K {{d_{il}^t}h_{il}^t{{\left( {{\bf{n}}_k^{\rm{q}}} \right)}^t}}  + {\bf{n}}_l^t$ is the effective noise distributed according to ${\mathcal{N}_\mathbb{C}}\left( {0,\sigma _{{\rm{eff}}}^2{{\bf{I}}_d}} \right)$ and $\sigma _{{\rm{eff}}}^2 = \sum\limits_{i = 1}^K {{d_{il}^t}\beta _{il}^t{\zeta _i}} \left( {1 - {\zeta _i}} \right){{\bf{I}}_d} + {\sigma ^2}{{\bf{I}}_d}$.

When the MRC scheme is employed, we set ${v_{kl}} = d_{kl}^t{h_{kl}}$, the local processed signal of ${\mathbf{s}}_{kl}^t$ at AP $l$ is given as
\begin{align}
&{\bf{\hat s}}_{kl}^t = v_{kl}^*{{\bf{y}}_l^t} = v_{kl}^*\sum\limits_{i = 1}^K {{h_{il}}{\bf{\mathord{\buildrel{\lower3pt\hbox{$\scriptscriptstyle\smile$}}
\over s} }}_i^t}  + v_{kl}^*{{\bf{n}}_l} \notag\\
&= d_{kl}^th_{kl}^*{h_{kl}}{\zeta _k}{\bf{s}}_k^t
+ d_{kl}^th_{kl}^*{h_{kl}}{\left( {{\bf{n}}_k^{\rm{q}}} \right)^t} + d_{kl}^th_{kl}^*\sum\limits_{i \ne k}^K {{h_{il}}{\bf{\mathord{\buildrel{\lower3pt\hbox{$\scriptscriptstyle\smile$}}
\over s} }}_i^t}  + d_{kl}^t\hat h_{kl}^*{{\bf{n}}_l}.
\end{align}
Then, the achievable SE for UE $k$ at communication round $t$ can be obtained in the following closed-form
\begin{equation}
{\rm{SINR}}_k^t = \frac{{p_k^tA_k^t}}{{\sum\limits_{i = 1}^K {p_i^tC_{ki}^t}  + \sum\limits_{i = 1}^K {p_i^tE_{ki}^t}  + F_k^t}},
\end{equation}
where
\begin{align}\label{SINR_DAC}
&A_k^t = {\left( {\zeta _{}^t} \right)^2}{\left| {\sum\limits_{l = 1}^L {d_{kl}^t{\beta _{kl}}} } \right|^2},\;\;\;C_{ki}^t = {\left( {\zeta _{}^t} \right)^2}\sum\limits_{l = 1}^L {d_{kl}^t{\beta _{kl}}} {\beta _{il}}, \notag\\
&E_k^t = d\sum\limits_{l = 1}^L {d_{il}^t{\beta _{kl}}{\beta _{il}}\zeta _{}^t\left( {1 - \zeta _{}^t} \right)},
\;\;\;F_k^t = d\sum\limits_{l = 1}^L {d_{kl}^t} {\beta _{kl}}{\sigma ^2}.
\end{align}
Therefore, the total uplink training time is
\begin{equation}
 T_{\text{time}}= \sum\limits_{t = 1}^T {\frac{S}{{R_k^t}}}  + \sum\limits_{t = 1}^T {\frac{{\left| {{{\cal K}^t}} \right|S}}{{\sum\limits_{k \in {\cal K}} {R_k^t} }}},
\end{equation}
where $\mathcal{K}^t = {\mathcal{D}_1^t} \cup {\mathcal{D}_2^t} \cdots {\mathcal{D}_L^t}$ and ${\mathcal{D}_l^t} = \left\{ {i:{d_{il}^t} = 1,i \in \left\{ {1, \cdots K} \right\}} \right\}$. Compared with (\ref{ut1}), it can be observed that choosing an appropriate serving UEs cluster ${{\cal D}_i^t}$ in each iteration can effectively reduce the total uplink training time.

\section{Differential Privacy Analysis }
DP is a privacy mechanism to fight against differential attacks and to ensure that the sensitive information of UEs is not exposed \cite{wahab2021federated}. The standard definition of DP imposes a point-wise upper bound on the divergence between the distributions $P\left( {\left. {\bf{y}} \right|{\cal B}} \right)$ and $P\left( {\left. {\bf{y}} \right|{\cal B}'} \right)$, where $\bf{y}$ is the received signal and ${\cal B}$ and ${\cal B}'$ are two ``neighboring'' global data sets which only differ by one sample at one UE.

In this section, we derive the upper-bound of the privacy preservation condition and provide the convergence analysis in the CF mMIMO-supported FL with noise injection by using low-resolution ADCs or DACs.

\subsection{Low-resolution ADCs Equipped at the APs}
\subsubsection{Privacy Preservation Condition}
\begin{myDef}
For two adjacent datasets ${{\mathcal{B}^{'}_j}} $ and ${{\mathcal{B}^{''}_j}}$ with ${\left| {{{\mathcal{B}^{'}_j}} - {{\mathcal{B}^{''}_j}}} \right|} = 1$ for some UEs $j$ and ${\left| {{{\mathcal{B}^{'}_i}} - {{\mathcal{B}^{''}_i}}} \right|} = 0$ for all $i \ne j$,
the communication and learning protocol is $({\mathbb{\epsilon}}, \delta)$-differentially private, where ${\mathbb{\epsilon}}>0$, and $\delta  \in \left[ {0,1} \right)$, when we have the following inequality for UE $k$ \cite{liu2020privacy}:
\begin{equation}\label{(14)}
\Pr\left( {\left. {{\mathbf{r}}_k^t} \right|{{\mathcal{B}^{'}_k}}} \right) \leqslant \exp \left(  {\mathbb{\epsilon}}\right)\Pr\left( {\left. {{\mathbf{r}}_k^t} \right|{{\mathcal{B}^{''}_k}}} \right) + \delta,
\end{equation}
where $\Pr$ refers to the probability of a certain event.
After $T$ iterations, the $({\mathbb{\epsilon}}, \delta)$ DP condition in (\ref{(14)}) can be written as
\begin{equation}\label{(18)}
\Pr \left( {\left| {\ln \left( {\prod\limits_{t = 1}^T {\frac{{P\left( {\left. {{\mathbf{r}}_k^t} \right|{\mathbf{r}}_k^{t - 1} \cdots {\mathbf{r}}_k^1,{\mathcal{B}_k^{'}}} \right)}}{{P\left( {\left. {{\mathbf{r}}_k^t} \right|{\mathbf{r}}_k^{t - 1} \cdots {\mathbf{r}}_k^1,{\mathcal{B}_k^{''}}} \right)}}} } \right)} \right| \!\leqslant\! {\mathbb{\epsilon}} } \right) \!\geqslant\! 1 - \delta .
\end{equation}
\end{myDef}

The $({\mathbb{\epsilon}}, \delta)$-DP condition ensures that for all possible adjacent datasets, the absolute value of the left side of (\ref{(18)}) can be bounded by ${\mathbb{\epsilon}}$ with probability at least $1-\delta$.
Note that the values ${\mathbb{\epsilon}}$ and $\delta$ stand for the similarity of the result distribution of the random mechanism performed on the data sets ${\mathcal{B}_k^{'}}$ and ${\mathcal{B}_k^{''}}$, and are interpreted as a privacy level \cite{dwork2014algorithmic}.
The lower ${\mathbb{\epsilon}}$ and $\delta$ indicate a higher level of privacy.

The sensitivity $\Delta _k^t$ of the noiseless received signal ${\mathbf{r}}_k^t - {\bm{\omega }}_k^t$ is defined as
\begin{align}\label{(19)}
\Delta _k^t &= \mathop {\max }\limits_{\mathcal{B}_k^{'},\mathcal{B}_k^{''}} \left\| {\alpha \sum\limits_{l = 1}^L {\sum\limits_{i = 1}^K {\sqrt {\frac{{p_i^t}}{{{{\left\| {{\mathbf{s}}_i^t\left( {\mathcal{B}_k^{'}} \right)} \right\|}^2}}}} h_{kl}^*{h_{il}}{\mathbf{s}}_i^t\left( {\mathcal{B}_k^{'}} \right)} }} \right. \notag\\
&\left.{- \alpha \sum\limits_{l = 1}^L {\sum\limits_{i = 1}^K {\sqrt {\frac{{p_i^t}}{{{{\left\| {{\mathbf{s}}_i^t\left( {\mathcal{B}_k^{''}} \right)} \right\|}^2}}}} h_{kl}^*{h_{il}}{\mathbf{s}}_i^t\left( {\mathcal{B}_k^{''}} \right)} } } \right\|.
\end{align}
Equation (\ref{(19)}) can be bounded as
\begin{equation}
\Delta _k^t \leqslant \mathop {\max }\limits_i 2\alpha \sqrt {p_i^t} \left| {h_{kl}^*{h_{il}}} \right|
\end{equation} with the help of \cite{wei2020federated} and the triangular inequality.
Then, according to (\ref{(10)}), we can obtain
\begin{equation}
{\ln \!\!\left( \!{\prod\limits_{t = 1}^T {\frac{{P\left( {\left. {{\mathbf{r}}_k^t} \right|{\mathbf{r}}_k^{t - 1} \cdots {\mathbf{r}}_k^1,{\mathcal{B}_k^{'}}} \right)}}{{P\left( {\left. {{\mathbf{r}}_k^t} \right|{\mathbf{r}}_k^{t - 1} \cdots {\mathbf{r}}_k^1,\mathcal{B}_k^{''}} \right)}}} } \right)}\!\!
=\!\! {\sum\limits_{t = 1}^T {\ln }\!\! \left( \!\!{\frac{{\exp \left( {\frac{{{{\left\| {{\bm{\omega }}_k^t} \right\|}^2}}}{{2{{\left( {m_k^t} \right)}^2}}}} \right)}}{{\exp \left( {\frac{{{{\left\| {{\bm{\omega }}_k^t + {\mathbf{v}}_k^t} \right\|}^2}}}{{2{{\left( {m_k^t} \right)}^2}}}} \right)}}} \right)},
\end{equation}
where
\begin{align}
{\mathbf{v}}_k^t &= \alpha \sum\limits_{l = 1}^L \sum\limits_{i = 1}^K \left( {\sqrt {\frac{{p_i^t}}{{{{\left\| {{\mathbf{s}}_i^t\left( {\mathcal{B}_k^{''}} \right)} \right\|}^2}}}} h_{kl}^*{h_{il}}{\mathbf{s}}_i^t\left( {\mathcal{B}_k^{''}} \right)}\right. \notag\\
&\left.{- \sqrt {\frac{{p_i^t}}{{{{\left\| {{\mathbf{s}}_i^t\left( {\mathcal{B}_k^{'}} \right)} \right\|}^2}}}} h_{kl}^*{h_{il}}{\mathbf{s}}_i^t\left( {\mathcal{B}_k^{'}} \right)} \right) ,
\end{align}
with $\left\| {{\mathbf{v}}_k^t} \right\| \leqslant \Delta _k^t$.

Following similar steps as in \cite[Appendix A]{dwork2014algorithmic}, we can then obtain the upper-bound on the privacy preservation condition
\begin{align}
\Pr \left( {\left| {\sum\limits_{t = 1}^T {\frac{{2{{\left( {{\bm{\omega }}_k^t} \right)}^T}{\mathbf{v}}_k^t + {{\left\| {{\mathbf{v}}_k^t} \right\|}^2}}}{{2{{\left( {m_k^t} \right)}^2}}}} } \right| > {\mathbb{\epsilon}} } \right)
&\mathop  \leqslant \limits^{\left( i \right)} \Pr \left( {\left| {\sum\limits_{t = 1}^T {\frac{{{{\left( {{\bm{\omega }}_k^t} \right)}^T}{\mathbf{v}}_k^t}}{{{{\left( {m_k^t} \right)}^2}}}} } \right| > {\mathbb{\epsilon}} - \sum\limits_{t = 1}^T {\frac{{{{\left\| {{\mathbf{v}}_k^t} \right\|}^2}}}{{2{{\left( {m_k^t} \right)}^2}}}} } \right)\notag\\
& = 2\Pr \left( {\sum\limits_{t = 1}^T {\frac{{{{\left( {{\bm{\omega }}_k^t} \right)}^T}{\mathbf{v}}_k^t}}{{{{\left( {m_k^t} \right)}^2}}}}  > {\mathbb{\epsilon}} - \sum\limits_{t = 1}^T {\frac{{{{\left\| {{\mathbf{v}}_k^t} \right\|}^2}}}{{2{{\left( {m_k^t} \right)}^2}}}} } \right)\notag\\
& \mathop  \leqslant \limits^{\left( {ii} \right)} 2\frac{1}{{\sqrt {2\pi \Lambda } }}\int_{ {\mathbb{\epsilon}}- \Lambda }^\infty  {\frac{x}{{ {\mathbb{\epsilon}}- \Lambda }}\exp } \left( { - \frac{{{x^2}}}{{2\Lambda }}} \right)dx,
\end{align}
where $(i)$ follows the inequality $\Pr \left( {\left| {X + a} \right| > {\mathbb{\epsilon}} } \right) \leqslant \Pr \left( {\left| X \right| + a > {\mathbb{\epsilon}} } \right)$ for an arbitrary $a \geqslant 0$, and $(ii)$ is due to
\begin{align}
\Pr \left( {\sum\limits_{t = 1}^T {\frac{{{{\left( {{\bm{\omega}} _k^t} \right)}^T}{\mathbf{v}}_k^t}}{{{{\left( {m_k^t} \right)}^2}}}}  > {\mathbb{\epsilon}} - {\Lambda} } \right)
&= \frac{1}{{\sqrt {2\pi \Lambda } }}\int_{ {\mathbb{\epsilon}}- \Lambda }^\infty  {\exp } \left( { - \frac{{{x^2}}}{{2\Lambda }}} \right)dx\notag\\
&\leqslant \frac{1}{{\sqrt {2\pi \Lambda } }}\int_{ {\mathbb{\epsilon}}- \Lambda }^\infty  {\frac{x}{{ {\mathbb{\epsilon}}- \Lambda }}\exp } \left( { - \frac{{{x^2}}}{{2\Lambda }}} \right)dx\notag\\
&= \frac{{\sqrt \Lambda  }}{{\sqrt {2\pi } \left( { {\mathbb{\epsilon}}- \Lambda } \right)}}\exp \left( { - \frac{{{{\left( { {\mathbb{\epsilon}}- \Lambda } \right)}^2}}}{{2\Lambda }}} \right),
\end{align}
where $\Lambda  \triangleq \sum\limits_{t = 1}^T {{{\left( {\frac{{\Delta _k^t}}{{m_k^t}}} \right)}^2}}$.
Finally, the closed-from $({\mathbb{\epsilon}}, \delta)$-DP condition is given by
\begin{equation}\label{(25)}
\frac{{\sqrt {2\Lambda } }}{{\sqrt \pi  \left( { {\mathbb{\epsilon}}- \Lambda } \right)}}\exp \left( { - \frac{{{{\left( { {\mathbb{\epsilon}}- \Lambda } \right)}^2}}}{{2\Lambda }}} \right) < \delta.
\end{equation}

\subsubsection{Convergence Analysis}
At the $t$-th iteration, the CPU estimates the scaled local gradient as ${\mathbf{r}}_k^t$, and then the global gradient is estimated as
\begin{align}
& \widehat {\nabla F}\left( {{{\mathbf{w}}^{t - 1}}} \right) = \frac{1}{{{B_{{\text{tot}}}}}}\sum\limits_{k = 1}^K {{\mathbf{r}}_k^{t - 1}}   = \alpha \nabla F\left( {{{\mathbf{w}}^{t - 1}}} \right) + \frac{\alpha }{{{B_{{\text{tot}}}}}}I^{t-1}
 + \frac{1}{{{B_{{\text{tot}}}}}}\sum\limits_{k = 1}^K {{\bm{\omega }}_k^{t - 1}},
\end{align}
where $I^{t-1} \triangleq \sum\limits_{k = 1}^K {\sqrt {\frac{{p_k^{t - 1}}}{{{{\left\| {{\mathbf{s}}_k^{t - 1}} \right\|}^2}}}} {\mathbf{s}}_k^{t - 1}} \sum\limits_{l = 1}^L {\sum\limits_{i \ne k}^K {{h_{kl}}h_{il}^*} }$.
With the help of \cite[Assumption 1]{liu2020privacy}, we have the following equality
\begin{align}
 F\left( {{{\mathbf{w}}^t}} \right)
&\leqslant \!F\left( {{{\mathbf{w}}^{t \!-\! 1}}} \right)\!\! + \!\!{\left[ {\nabla F\left( {{{\mathbf{w}}^{t \!-\! 1}}} \right)} \right]^T}\left[ {{{\mathbf{w}}^t} \!\!-\!\! {{\mathbf{w}}^{t \!-\! 1}}} \right] \!\!+\!\!  \frac{M}{2}{\left\| {{{\mathbf{w}}^t} \!- \! {{\mathbf{w}}^{t\! -\! 1}}} \right\|^2}\notag\\
& = F\left( {{{\mathbf{w}}^{t - 1}}} \right) \!-\! {\left[ {\nabla F\left( {{{\mathbf{w}}^{t - 1}}} \right)} \right]^T}\left[ {\alpha \eta \nabla F\left( {{{\mathbf{w}}^{t - 1}}} \right) + \frac{{\alpha \eta }}{{{B_{{\text{tot}}}}}}{I^{t - 1}} + \frac{\eta }{{{B_{{\text{tot}}}}}}\sum\limits_{k = 1}^K {{\bm{\omega }}_k^{t - 1}} } \right] \notag\\
&+ \frac{M}{2}\left\| {\alpha \eta \nabla F\left( {{{\mathbf{w}}^{t - 1}}} \right) + \frac{{\alpha \eta }}{{{B_{{\text{tot}}}}}}{I^{t - 1}}} \right. {\left. { + \frac{\eta }{{{B_{{\text{tot}}}}}}\sum\limits_{k = 1}^K {{\bm{\omega }}_k^{t - 1}} } \right\|^2},
\end{align}
where $M$ is a positive constant.
Setting $\eta  = \frac{1}{M}$ and taking expectation over the randomness of additive noise, we have
\begin{align}
\mathbb{E}\left\{ {F\left( {{{\mathbf{w}}^t}} \right)} \right\} & \leqslant F\left( {{{\mathbf{w}}^{t - 1}}} \right) - \frac{\alpha }{M}{\left\| {\nabla F\left( {{{\mathbf{w}}^{t - 1}}} \right)} \right\|^2}\!\! -\!\! {\left[ {\nabla F\left( {{{\mathbf{w}}^{t - 1}}} \right)} \right]^T} {\frac{\alpha }{{M{B_{{\text{tot}}}}}}{I^{t - 1}}} \notag\\
& + \frac{{{\alpha ^2}}}{{2M}}{\left\| {\nabla F\left( {{{\mathbf{w}}^{t - 1}}} \right)} \right\|^2} + \left( {{{\left[ {\nabla F\left( {{{\mathbf{w}}^{t - 1}}} \right)} \right]}^T}\frac{{{\alpha ^2}}}{{M{B_{{\text{tot}}}}}}{I^{t - 1}}} \right)\notag\\
& + \frac{1}{{2M}}{\left\| {\frac{\alpha }{{{B_{{\text{tot}}}}}}{I^{t - 1}}} \right\|^2} + \frac{1}{{2M}}\frac{d}{{B_{{\text{tot}}}^2}}\sum\limits_{k = 1}^K {{{\left( {m_k^t} \right)}^2}}.
\end{align}
According to \cite[Assumption 2]{liu2020privacy}, we obtain
\begin{align}
\mathbb{E}\left\{ {F\left( {{{\mathbf{w}}^t}} \right)} \right\} - {F^*}
&\leqslant \frac{{M - \alpha \left( {2 - \alpha } \right)\mu }}{M}\left( {\mathbb{E}\left\{ {F\left( {{{\mathbf{w}}^{t - 1}}} \right)} \right\} - {F^*}} \right) \notag\\
&+ \frac{{{\alpha ^2}}}{{2MB_{{\text{tot}}}^2}}{\left\| {\frac{\alpha }{{{B_{{\text{tot}}}}}}{I^{t - 1}}} \right\|^2} + \frac{1}{{2M}}\frac{d}{{B_{{\text{tot}}}^2}}\sum\limits_{k = 1}^K {{{\left( {m_k^{t-1}} \right)}^2}},
\end{align}
where $\mu$ is a positive constant. Therefore, by applying the above inequality repeatedly through $T$ iterations, the results follow immediately.
Finally, the average optimality gap after $T$ iteration is upper bounded by
\begin{align}\label{conver}
\mathbb{E}\left\{ {F\left( {{{\mathbf{w}}^t}} \right)} \right\} - {F^*}
&\leqslant {\left( {1 - \frac{{\alpha \left( {2 - \alpha } \right)\mu }}{M}} \right)^T}\left( {\mathbb{E}\left\{ {F\left( {{{\mathbf{w}}^{1}}} \right)} \right\} - {F^*}} \right) \notag\\
&+ \frac{{{\alpha ^2}}}{{2MB_{{\text{tot}}}^2}}\sum\limits_{t = 1}^T {{{\left( {1 - \frac{{\alpha \left( {2 - \alpha } \right)\mu }}{M}} \right)}^{T - t}}}{\left\| {\frac{\alpha }{{{B_{{\text{tot}}}}}}{I^{t}}} \right\|^2} \notag\\
&+ \frac{1}{{2M}}\frac{d}{{B_{{\text{tot}}}^2}}\sum\limits_{t = 1}^T {{{\left( {1 - \frac{{\alpha \left( {2 - \alpha } \right)\mu }}{M}} \right)}^{T - t}}}\sum\limits_{k = 1}^K {{{\left( {m_k^t} \right)}^2}}.
\end{align}
The first item in (\ref{conver}) reflects the initial optimality gap $\left( {\mathbb{E}\left\{ {F\left( {{{\mathbf{w}}^{1}}} \right)} \right\} - {F^*}} \right)$ increases with the increase of $T$, and the third considers the effect of effective additional noise power. Interestingly, the bound in (\ref{conver}) indicates that the quantization noise added in the initial iteration would enlarge the final optimality gap less than the noise added in the later iterations. This is because the contribution of the noise added in the iteration $t$ is affected by a factor ${{{\left( {1 - \frac{{\alpha \left( {2 - \alpha } \right)\mu }}{M}} \right)}^{T - t}}}$.

\subsection{Low-resolution DACs at the UEs}
In this case, the privacy preservation condition becomes
\begin{align}\label{1}
{{\cal L}_{{\cal B},{\cal B}'}}\left( {{\bf{y}}_l^t} \right)
= \ln \left( {\prod\limits_{t = 1}^T {\frac{{P\left( {\left. {{\bf{y}}_l^t} \right|{\bf{y}}_l^{t - 1}, \cdots {\bf{y}}_l^1,{\cal B}} \right)}}{{P\left( {\left. {{\bf{y}}_l^t} \right|{\bf{y}}_l^{t - 1}, \cdots {\bf{y}}_l^1,{\cal B}'} \right)}}} } \right)
= \sum\limits_{t = 1}^T {\ln } \left( {\frac{{P\left( {\left. {{\bf{y}}_l^t} \right|{\bf{y}}_l^{t - 1}, \cdots {\bf{y}}_l^1,{\cal B}} \right)}}{{P\left( {\left. {{\bf{y}}_l^t} \right|{\bf{y}}_l^{t - 1}, \cdots {\bf{y}}_l^1,{\cal B}'} \right)}}} \right),
\end{align}
where
\begin{align}
P\left( {\left. {{\bf{y}}_l^t} \right|{\bf{y}}_l^{t - 1}, \cdots {\bf{y}}_l^1,{\cal B}} \right)
=\frac{1}{{\sigma _{{\rm{eff}}}^2\sqrt {2\pi } }}{\rm{exp}}\left( { - \frac{{{{\left\| {{\bf{y}}_l^t - \sum\limits_{i = 1}^K {{\alpha _i}d_{il}^th_{il}^t{\bf{x}}_i^t\left( {{{\cal B}_i}} \right)} } \right\|}^2}}}{{2{{\left( {\sigma _{{\rm{eff}}}^2} \right)}^2}}}} \right),
\end{align}
Then, (\ref{1}) can be rewritten as
\begin{align}
&{{\cal L}_{{\cal B},{\cal B}'}}\left( {{\bf{y}}_l^t} \right)
= \sum\limits_{t = 1}^T {\ln } \left( {{{{\rm{exp}}\left( { - \frac{{{{\left\| {{\bf{y}}_l^t - \sum\limits_{i = 1}^K {{\alpha ^t}h_{il}^t{\bf{s}}_i^t\left( {{{\cal B}_i}} \right)} } \right\|}^2}}}{{2{{\left( {\sigma _{{\rm{eff}}}^2} \right)}^2}}}} \right)}}/{{{\rm{exp}}\left( { - \frac{{{{\left\| {{\bf{y}}_l^t - \sum\limits_{i = 1}^K {{\alpha ^t}h_{il}^t{\bf{s}}_i^t\left( {{{{\cal B}'}_i}} \right)} } \right\|}^2}}}{{2{{\left( {\sigma _{{\rm{eff}}}^2} \right)}^2}}}} \right)}}} \right)\notag\\
&= \sum\limits_{t = 1}^T {\ln } \left( {{{{\rm{exp}}\left( { - \frac{{{{\left\| {{\bm{\omega }} _l^t} \right\|}^2}}}{{2{{\left( {\sigma _{{\rm{eff}}}^2} \right)}^2}}}} \right)}}/{{{\rm{exp}}\left( { - \frac{{{{\left\| {{\bm{\omega }} _l^t + {\bf{v}}_l^t} \right\|}^2}}}{{2{{\left( {\sigma _{{\rm{eff}}}^2} \right)}^2}}}} \right)}}} \right)= \sum\limits_{t = 1}^T {\frac{{{{\left\| {{\bf{v}}_l^t} \right\|}^2} + 2{{\left( {{\bm{\omega }} _l^t} \right)}^T}{\bf{v}}_l^t}}{{2{{\left( {\sigma _{{\rm{eff}}}^2} \right)}^2}}}}=\Gamma _l^t.
\end{align}
Since the sensitivity $\Delta _l^t$ is given as
\begin{align}
{\bf{v}}_l^t = \sum\limits_{i = 1}^K {{\alpha _i}\left| {h_{il}^t} \right|} \left( {{\bf{x}}_i^t\left( {{{{\cal B}'}_i}} \right) - {\bf{x}}_i^t\left( {{{\cal B}_i}} \right)} \right),
\left\| {{\bf{v}}_l^t} \right\| \le 2\mathop {\max }\limits_i \sqrt {p_i^t} \left| {h_{il}^t} \right| = \Delta _l^t,
\end{align}
the upper-bound on the privacy preservation condition can be derived as
\begin{align}
\Pr \left( {\left| {\Gamma _l^t} \right| > {\mathbb{\epsilon}}} \right)
&\le \Pr \left( {\left| {\sum\limits_{t = 1}^T {\frac{{{{\left( {{\bm{\omega}} _l^t} \right)}^T}{\bf{v}}_l^t}}{{{{\left( {\sigma _{{\rm{eff}}}^2} \right)}^2}}}} } \right| >  {\mathbb{\epsilon}}- \sum\limits_{t = 1}^T {\frac{{{{\left\| {{\bf{v}}_l^t} \right\|}^2}}}{{2{{\left( {\sigma _{{\rm{eff}}}^2} \right)}^2}}}} } \right)\notag\\
& = 2\Pr \left( {\sum\limits_{t = 1}^T {\frac{{{{\left( {{\bm{\omega}} _l^t} \right)}^T}{\bf{v}}_l^t}}{{{{\left( {\sigma _{{\rm{eff}}}^2} \right)}^2}}}} } \right) >  {\mathbb{\epsilon}}- \sum\limits_{t = 1}^T {\frac{{{{\left\| {{\bf{v}}_l^t} \right\|}^2}}}{{2{{\left( {\sigma _{{\rm{eff}}}^2} \right)}^2}}}} \notag\\
& \le 2\Pr \left( {\Lambda  >  {\mathbb{\epsilon}}- \sum\limits_{t = 1}^T {\frac{1}{2}{{\left( {\frac{{\Delta _l^t}}{{\sigma _{{\rm{eff}}}^2}}} \right)}^2}} } \right)\notag\\
& = \frac{2}{{\sqrt {2\pi \sum\limits_{t = 1}^T {{{\left( {\frac{{\Delta _l^t}}{{\sigma _{{\rm{eff}}}^2}}} \right)}^2}} } }}\int_b^\infty  {\exp \left( { - \frac{{{x^2}}}{{2\sum\limits_{t = 1}^T {{{\left( {\frac{{\Delta _l^t}}{{\sigma _{{\rm{eff}}}^2}}} \right)}^2}} }}} \right)} dx,
\end{align}
where
\begin{align}
\begin{array}{l}
b =  - \sum\limits_{t = 1}^T {{{\left( {\frac{{\Delta _l^t}}{{{\sigma _{{\rm{eff}}}}}}} \right)}^2}}  > 0,\;\;\;
\frac{2}{{\sqrt {2\pi \nu } }}\int_{ - \frac{\nu }{2}}^\infty  {\exp \left( { - \frac{{{x^2}}}{{2\nu }}} \right)} dx < \delta, \;\;\;
\nu  = \sum\limits_{t = 1}^T {{{\left( {\frac{{\Delta _l^t}}{{{\sigma _{{\rm{eff}}}}}}} \right)}^2}}.
\end{array}
\end{align}
Besides, the convergence can be proved following similar steps as in Section III-A-(2).

\section{Power Control for Training Time Minimization and An asynchronous FL protocol}
In this section, an optimization problem is formulated to jointly optimize the transmit power and data rate under the practical constraints on UEs¡¯ energy consumption with different quantization accuracies of the ADCs/DACs. Note that the quantization noise is used for privacy protection, therefore, different quantization accuracies realizes different DP protection levels. By applying the successive convex approximation approach, we design a computationally-efficient algorithm to obtain a suboptimal solution of the power allocation. Besides, we propose an asynchronous FL protocol to alleviate the staleness, efficiency and better utilize the progress made by stragglers.

\subsection{Low-resolution ADCs Equipped at the APs}
According to (\ref{R}) and (\ref{(16)}), the problem of FL training time minimization with low-resolution ADCs in the CF mMIMO system can be formulated as
\begin{align}
& \mathop {\text{minimize} }\limits_{{\mathbf{p}},{\mathbf{R}}}\;\;\;T_{\text{time}}\label{P1}\tag{47a}\\
& \;\;{\text{s.}}{\text{t.}}\;\;\;0 \leqslant {p_k} \leqslant {p_{\max }},\tag{47b}\\
& \;\;\;\;\;\;\;\;{R_k} \leqslant {r_k},\tag{47c}
\end{align}
where ${\mathbf{p}} = {\left[ {{p_1}, \cdots ,{p_K}} \right]^T},{\mathbf{R}} = {\left[ {{R_1}, \cdots ,{R_K}} \right]^T}$.
By introducing slack variables ${u_k}$, $x$, $x_1$, and $x_2$,  we reformulate (\ref{P1}) as follows:
\textcolor{black}{
\begin{align}\label{P2}
& \mathop {\text{minimize} }\limits_{{\mathbf{u}},{\mathbf{R}},x}\;\;\; x \tag{48a}\\
& \;{\text{s.}}{\text{t.}}\;\;\;x \geqslant {x_1} + {x_2}, \tag{48b}\\
& \;\;\;\;\;\;\;\;{x_1} \geqslant \frac{ST}{{{R_k}}}, \notag
 \;\;\;\;\;\;\;\;{x_2} \geqslant \frac{{KST}}{{\sum\limits_{k = 1}^K {{R_k}} }}, \tag{48c}\\
& \;\;\;\;\;\;\;\;{R_k} \leqslant \left( {1 - \frac{{{\tau _p}}}{{{\tau _c}}}} \right)B \times {\log _2}\left( {1 + \frac{{u_k^2{A_k}}}{{\sum\limits_{i \ne k}^L {u_i^2} {B_{ki}} + {C_k} +  {D_k} + \sum\limits_{i = 1}^K {u_i^2} {E_{ki}}}}} \right), \label{(37)}\tag{48d}\\
& \;\;\;\;\;\;\;\;u_k^2 \leqslant {p_{\max }},
 \;\;\;\;\;\;\;\;{u_k} \geqslant 0.\tag{48e}
\end{align}}
Note that different quantization accuracies realize different DP protection levels. Specifically, in order to satisfy the $({\mathbb{\epsilon}}, \delta)$-differentially private, according to (\ref{(25)}), the appropriate value of the linear gain $\alpha$ which depends on the number of quantization bits needs to be selected. At the same time, according to (\ref{R_qu}), the rate $R_k$ is affected by $\alpha$.
However, due to the nonconvex constraint (\ref{(37)}), (55) is still challenging. To address the problem at hand, we exploit the fact that a function $f\left( {x,y} \right) = \log_2 \left( {1 + \frac{{{{\left| x \right|}^2}}}{y}} \right)$ has the following lower bound \cite{vu2020cell}:
\begin{align}
f\left( {x,y} \right) & \geqslant \log_2 \left( {1 + \frac{{{{\left| {{x^{\left( n \right)}}} \right|}^2}}}{{{y^{\left( n \right)}}}}} \right) - \frac{{{{\left| {{x^{\left( n \right)}}} \right|}^2}}}{{{y^{\left( n \right)}}}} + 2\frac{{{x^{\left( n \right)}}x}}{{{y^{\left( n \right)}}}} - \frac{{{{\left| {{x^{\left( n \right)}}} \right|}^2}\left( {{{\left| x \right|}^2} + y} \right)}}{{{y^{\left( n \right)}}\left( {{{\left| {{x^{\left( n \right)}}} \right|}^2} + {y^{\left( n \right)}}} \right)}},\tag{49}
\end{align}
where $n$ denotes the $n$-th iteration of SCA, $x \in \mathbb{R},y > 0$, and ${y^{\left( n \right)}} > 0$. Therefore, the concave lower bound in (\ref{(37)}) is
\begin{align}
& {R_k} \leqslant \left( {1 - \frac{{{\tau _p}}}{{{\tau _c}}}} \right)B\left( {{{\log }_2}\left( {1 + \frac{{{{\left| {\Upsilon _k^{\left( n \right)}} \right|}^2}}}{{\Pi _k^{\left( n \right)}}}} \right) - \frac{{{{\left| {\Upsilon _k^{\left( n \right)}} \right|}^2}}}{{\Pi _k^{\left( n \right)}}} + 2\frac{{\Upsilon _k^{\left( n \right)}{\Upsilon _k}}}{{\Pi _k^{\left( n \right)}}} - \frac{{{{\left| {\Upsilon _k^{\left( n \right)}} \right|}^2}\left( {\Upsilon _k^2 + {\Pi _k}} \right)}}{{\Pi _k^{\left( n \right)}\left( {{{\left| {\Upsilon _k^{\left( n \right)}} \right|}^2} + \Pi _k^{\left( n \right)}} \right)}}} \right),\tag{50}
\end{align}
where ${\Upsilon _k} = {u_k}\sqrt {{A_k}}$, and
${\Pi _k} =\sum\limits_{i \ne k}^L {u_i^2} {B_{ki}} + {C_k} + {D_k} +\sum\limits_{i = 1}^K {{p_i}} {E_{ki}}$.

Finally, at iteration $t$, problem (48) can be approximated by the following convex problem for given point $u_k^{\left( n \right)}$:
\begin{equation}\label{P3}\tag{51}
\mathop {\min }\limits_{\left\{ {{\mathbf{u}},{\mathbf{R}}} \right\} \in \mathcal{F}} x,
\end{equation}
where $\mathcal{F} \triangleq \left\{ { \text{(48b), (48c), (48e), (50)}} \right\}$ is a convex feasible set. As a result, it can be solved by convex optimization. Besides, we can further tighten the bounds in (48) iteratively, which is suboptimal in Algorithm 1. Note that the proposed algorithm can achieve a suboptimal solution of (48) and the corresponding convergence can be proved via a similar approach in \cite{vu2020cell}.
\textcolor{black}{It is noticed that problem (51) is solving simple convex programs. Therefore, the complexity of Algorithm 1 to solve the problem (51) in the suboptimal method is in polynomial time.}

\begin{algorithm}[t]
\caption{A Suboptimal Algorithm for (48)}
\label{algorithm}
\begin{algorithmic}[1]
\Require Set the maximum transmit power for each UE as ${p_{\max }}$; Large-scale fading coefficients ${\beta _{lk}},\forall l,k$. Initial values for $u_k^{\left( 0 \right)},\forall k$, and the tolerance $\varepsilon  \ge 0$. Set up $n=1$.
\Ensure The optimal solutions $u _k^{{\rm{opt}}} = u _k^{\left( n \right)},R_k^{{\rm{opt}}} = R_k^{\left( n \right)},\;\forall k$.
\State Iteration $n$:
\begin{itemize}
  \item Solve (\ref{P3}) to get its optimal solution $\left\{ {{{\mathbf{u}}^*},{{\mathbf{R}}^*}} \right\}$
  \item Update $\left\{ {{{\mathbf{u}}^{\left( n \right)}},{{\mathbf{R}}^{\left( n \right)}}} \right\} = \left\{ {{{\mathbf{u}}^*},{{\mathbf{R}}^*}} \right\}$.
\end{itemize}
\label{Step_1}
\State Stop if $\left| {x - {x^{\left( n \right)}}} \right| \leqslant \varepsilon$. Otherwise, go to Step \ref{Step_3}.
\State Set $n=n+1$, then go to Step \ref{Step_1}.
\label{Step_3}
\end{algorithmic}
\end{algorithm}

\subsection{Low-resolution DACs at the UEs}
The problem of FL training time minimization with low-resolution DACs in the CF mMIMO system can be formulated as
\textcolor{black}{\begin{align}
&\mathop {\min }\limits_{{\bf{R}},{\bf{u}}} \;\;\;w\tag{52a}\\
&\;{\rm{s.}}{\rm{t.}}\;\;\;\;w \ge {\rm{ }}{t_1} + {t_2},\tag{52b}\\
&\;\;\;\;\;\;\;\;\;\;{t_1} \ge \sum\limits_{t = 1}^T {\frac{{{S_{u,k}}}}{{R_k^t}}}, \;\;\;{t_2} \ge \sum\limits_{t = 1}^T {\frac{{\left| {{{\cal K}^t}} \right|{S_u}}}{{\sum\limits_{k \in {\cal K}} {R_k^t} }}}, \tag{52c}\\
&\;\;\;\;\;\;\;\;\;\;R_k^t \le \left( {1 - \frac{{{\tau _p}}}{{{\tau _c}}}} \right)B{\log _2}\left( {1 + {\rm{SINR}}_k^t} \right),\tag{52d}\\
&\;\;\;\;\;\;\;\;\;{\left( {u_k^t} \right)^2} \le {p_{\max }},\tag{52e}
\end{align}}
where
\begin{equation}
{\rm{SINR}}_k^t = \frac{{p_k^tA_k^t}}{{\sum\limits_{i = 1}^K {p_i^tB_{ki}^t}  + \sum\limits_{i = 1}^K {p_i^tC_{ki}^t}  + E_k^t}},\tag{53}
\end{equation}
and $A_k^t$, $B_{ki}^t$, $C_{ki}^t$, and $E_k^t$ are defined in (\ref{SINR_DAC}).
Note that problem (52) can be solved follows the similar steps to problem (48).

\subsection{An Asynchronous FL Protocol}
\textcolor{black}{Synchronous FL considers synchronous communication during training round between the CPU and UEs. For the former, the CPU should wait until getting response from sufficient UEs. Unfortunately, some UEs may be unresponsive in the training process due to vulnerable wireless network. Then, the CPU drops such UEs and proceeds on to the next training process. On the contrary, asynchronous FL enables all UEs to directly send update information to the CPU after every local round that is dropped in synchronous FL optimization. In general, asynchronous FL can achieve faster convergence when wireless links are vulnerable and heterogeneous across the UEs. Therefore, asynchronous FL has drawn more interests in recent papers.}

It can be observed from problem (52) that $d_{kl}$ would determine the required training time. Therefore, we propose a simple asynchronous FL protocol based on the large-scale fading coefficient. Specifically, a lag-tolerant algorithm which allows some UEs to stay asynchronous with the CPU is provided.
Note that straggler UEs refer to UEs that are slower and are still training locally based on outdated models. Generally, the UE should start training based on the latest global model received from the CPU.

At the communication iteration $t$, each AP classifies all UEs into three categories: synchronous, asynchronous and need-to-be-synchronized, respectively.
First, the synchronous UEs refers to the UEs that are served by at least one AP and \textcolor{black}{complete} its local model update based on the latest global model transmitted from the CPU.
Then, since the channel gains of straggler UEs are not strong enough, they may dramatically slow down the whole FL process. In order to mitigate the straggler effect, the local training of straggler UEs are still based on the last version of the global model. Therefore, the straggler UEs are also called asynchronous UEs.
Besides, we assume that all the UEs need to update their local training model at least are forced to synchronize after at most $T_{\text{tol}}$ blocks so that the global model will not be poisoned by the seriously outdated local models, where $T_{\text{tol}}$ is called the lag tolerance.

In this model, at iteration $t$, the $l$th AP only serves $K_{0,l} \le K$ APs corresponding to the $K_{0,l}$ largest large-scale fading coefficients. The main question arising immediately is how to choose $K_{0,l}$. Naturally, we can choose $K_{0,l}$ UEs satisfying
\begin{equation}
\sum\limits_{i = 1}^{{K_{0,l}}} {\frac{{{{\bar \beta }_{il}}}}{{\sum\limits_{i' = 1}^K {{\beta _{i'l}}} }}}  \ge \nu \%,\tag{54}
\end{equation}
where $\left\{ {{{\bar \beta }_{1l}}, \cdots ,{{\bar \beta }_{Kl}}} \right\}$ is the sorted (in descending order) set of the set $\left\{ {{\beta _{1l}}, \cdots ,{\beta _{Kl}}} \right\}$, and $\nu$ is the lag percent. After choosing ${\cal D}_1^t, \cdots ,{\cal D}_L^t$, where ${\mathcal{D}_l^t} = \left\{ {i:{d_{il}^t} = 1,i \in \left\{ {1, \cdots K} \right\}} \right\}$, we can follow the same method as in Section IV-A to an optimized power control.

\section{Numerical Results}
In this section, we first evaluate the performance of the proposed power control algorithm under different DP protection levels in the CF mMIMO supported FL network. Note that the DP mechanism is realized by the quantization noise. Therefore, we adopt ADCs and DACs with different quantization bits to achieve different DP protection levels.
\textcolor{black}{More specifically}, we adopt similar parameters setting as in \cite{vu2020cell}.
Note that we do not propose a new FL framework but rather apply a existed FL framework in CF mMIMO systems. We focus on how to reduce the training time using different quantization bits or under differential privacy levels. Therefore, the simulation on real datasets to see the effectiveness of the considered FL framework has already performed in \cite{liu2020privacy} and hence, they are not considered in this paper. \textcolor{black}{Note that the unit of time is seconds in the following figures.}

\begin{figure}[t!]
\centering
\includegraphics[scale=0.7]{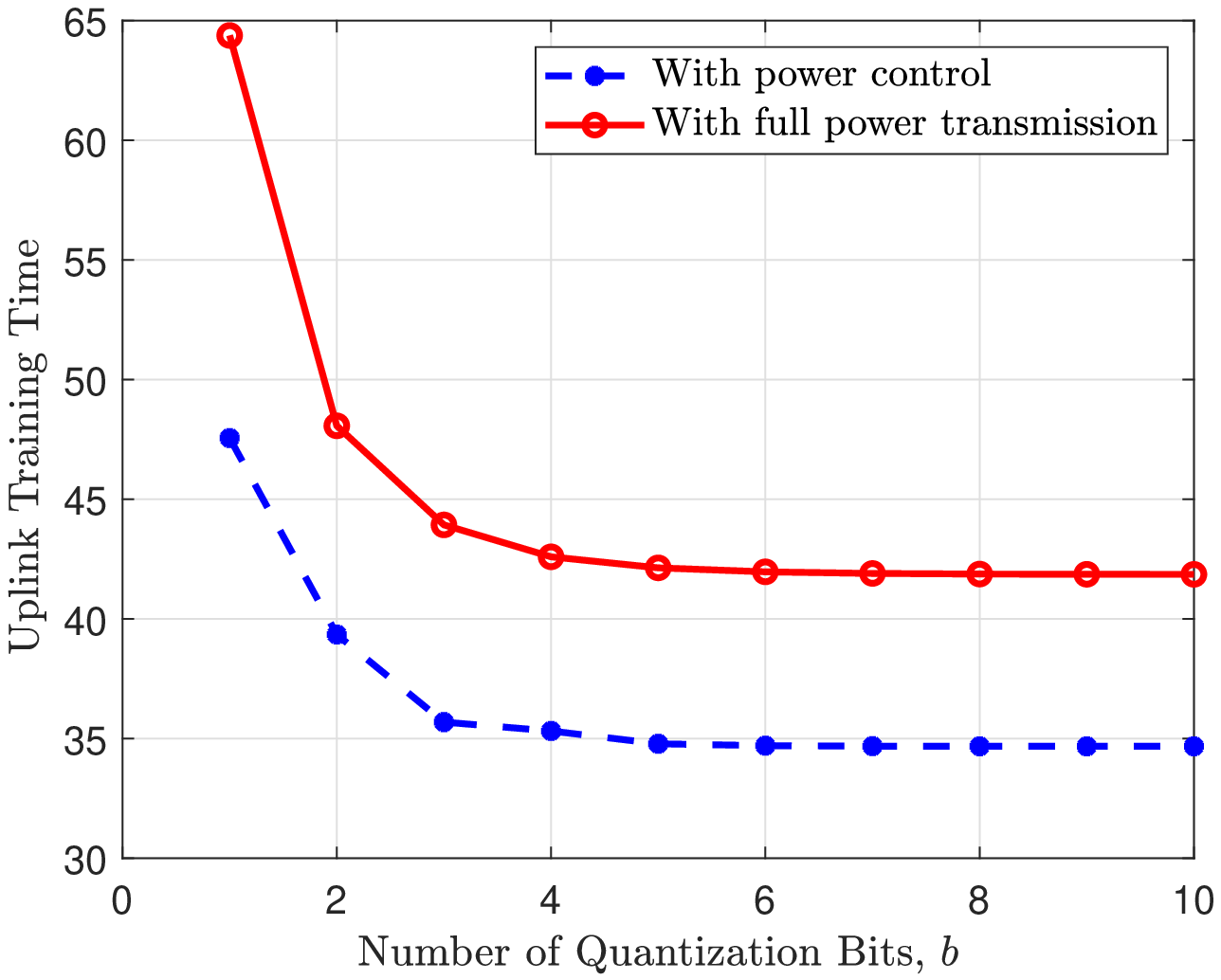}
\caption{Uplink training time against the number of quantization bits with $L=10$, $K =3$, $D=1$ km, and $p_{\max} = 200$ mW.}
\label{fig2}
\end{figure}

Fig. \ref{fig2} evaluates the uplink training time as a function of the number of quantization bits in the synchronized model. We also compare the performance of applying the proposed power control method and the full power transmission with the synchronized model. It can be seen from Fig. \ref{fig2} that the uplink training time decreases as the number of quantization bits increase since the quantization error reduces. Specifically, the performance of $b=3$ is close to that of $b=10$, which we refer to as the perfect ADCs case. Therefore, reducing the quantization bits from 10 to 3 only has a slight impact on the performance of uplink training time. Also, the fronthaul load can be relaxed significantly since the data size is reduced. Furthermore, applying the proposed power control method leads to a huge reduction in terms of uplink training time. In particular, the uplink training time reduces 35\% and 21\% when $b=1$ and $b=10$, respectively.

\begin{figure}[t!]
\centering
\includegraphics[scale=0.7]{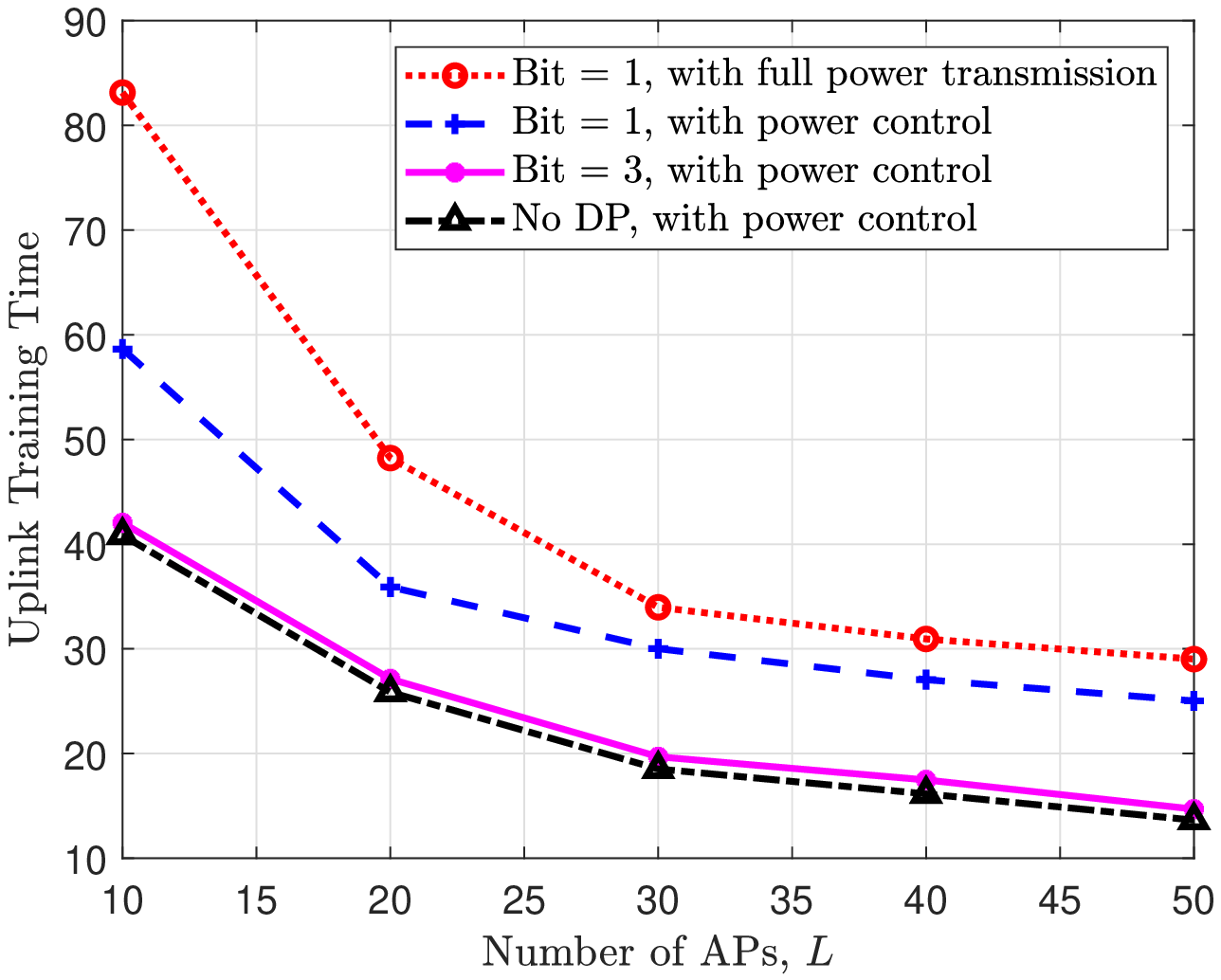}
\caption{Uplink training time against the number of APs with $K =3$, $D=1$ km, and $p_{\max} = 200$ mW.}
\label{fig3}
\end{figure}

Fig. \ref{fig3} shows the uplink training time against the number of APs in the synchronized model. As expected, the uplink training time decreases along with the increase of the number of APs, which is resulted from from the higher macro-diversity gain for enhancing the detection performance. Besides, when $b=1$, applying the proposed power control method leads to a 42\% reduction in terms of uplink training time compared with the one with simple full power transmission. Although the performance gap between the proposed power control and the full power transmission reduces when the number of APs increases, the advantage of our power control method is still obvious for a reasonable number of APs. Furthermore, the performance with $b=3$ approaches that of perfect ADCs. Note that the latter means the quantization error is not dominated and hence, the DP also no longer exists. This indicates that one can use low-resolution ADCs to realize privacy-preserving with reduced hardware cost and fronthaul load without increasing the uplink training time noticeably.

\begin{figure}[t!]
\centering
\includegraphics[scale=0.7]{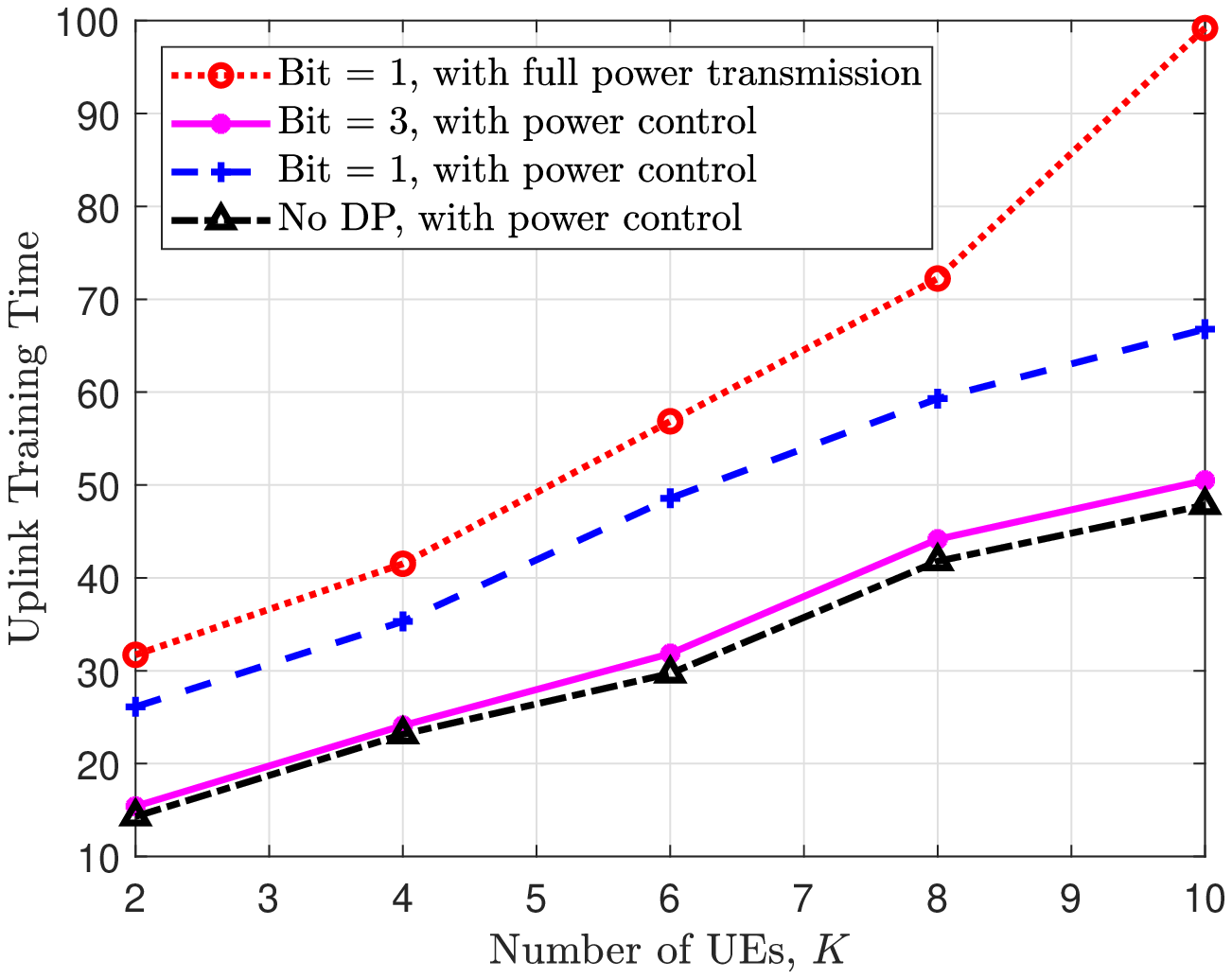}
\caption{Uplink training time against the number of UEs with $L=30$, $D=1$ km, and $p_{\max} = 200$ mW.}
\label{fig4}
\end{figure}

Fig. \ref{fig4} compares the uplink training times as a function of the number of UEs in the synchronized model. As can be seen, the performance of uplink training time becomes longer when the system has more UEs. This is because the mutual interference becomes stronger for a larger number of UEs. However, our proposed power control method can effectively mitigate the impact of mutual interference.
In particular, the performance of uplink training time is effectively reduced by 48\%. \textcolor{black}{Moreover}, the performance gap between $b=3$ and $b=10$ is also negligible, although the gaps are enlarged with the increases of the number of UEs. For example, when $K=10$, taking $b=3$ leads to 5\% performance loss.

\begin{figure}[t!]
\centering
\includegraphics[scale=0.7]{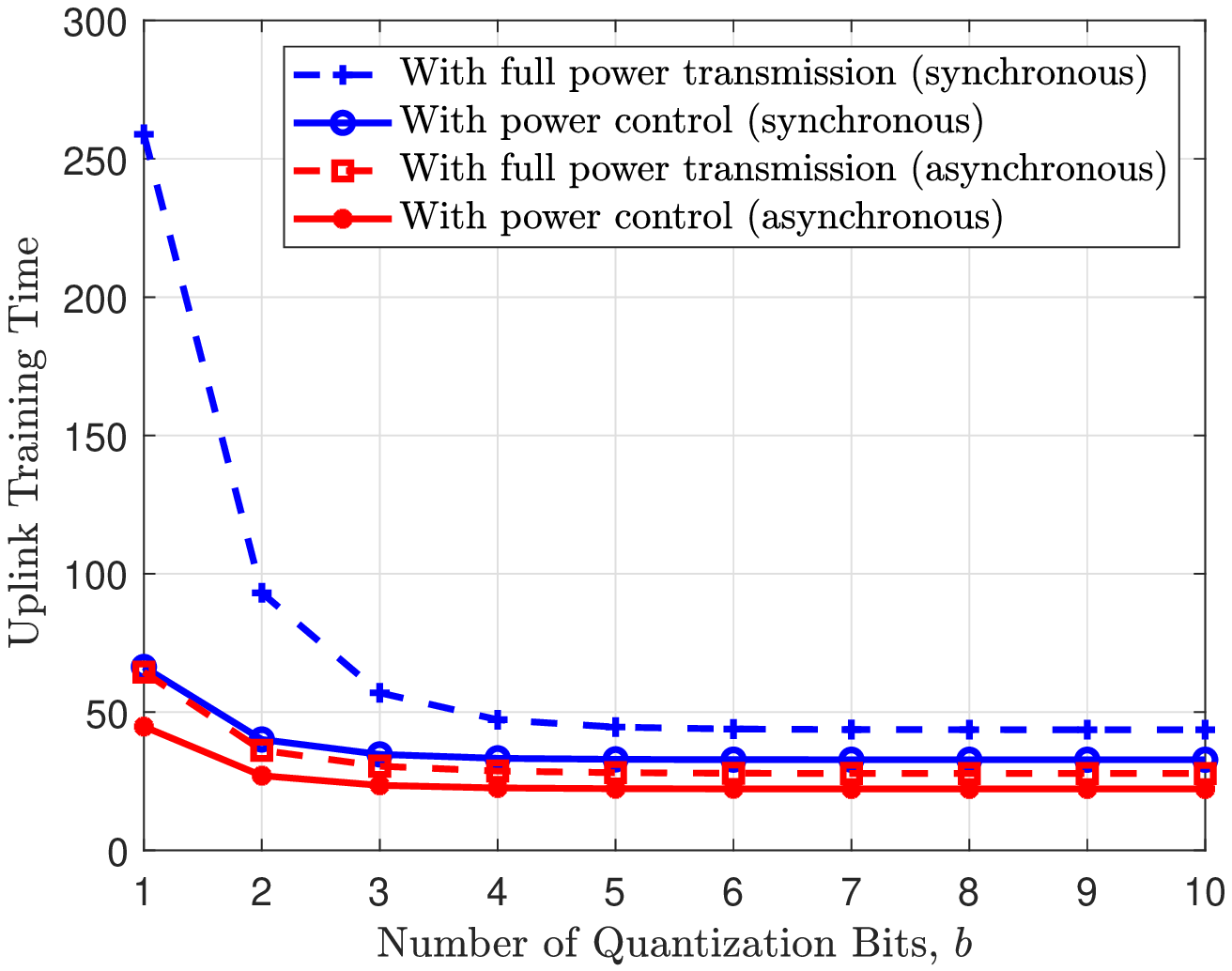}
\caption{Uplink training time against the number of quantization bits with $L = 10$, $K =4$, $D=1$ km, and $p_{\max} = 200$ mW.}
\label{fig5}
\end{figure}

Fig. \ref{fig5} \textcolor{black}{presents} the performance of synchronous mode and synchronous mode as a function of the number of quantization bits when low-resolution DACs are equipped at the UEs. It can be observed that utilizing the asynchronous mode can effectively reduce the training time. By using the asynchronous mode, we can decrease the number of UEs that each AP needs to be served. Consequently, the uplink training time reduces sharply. In particular, when $b=1$, the uplink training time with the asynchronous mode and power control is 2.5 times shorter than that of the synchronous mode.

\begin{figure}[t!]
\centering
\includegraphics[scale=0.7]{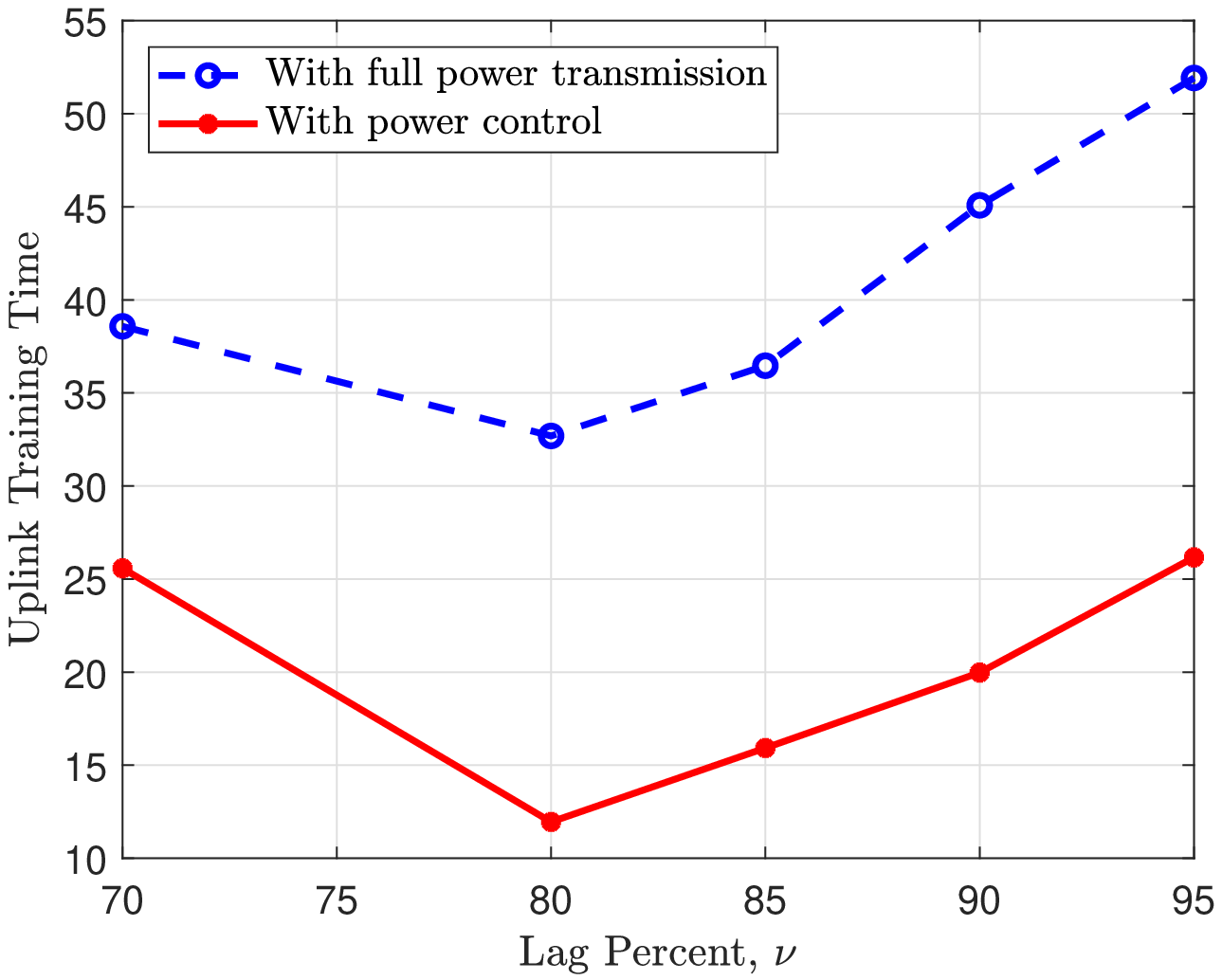}
\caption{Uplink training time against the lag tolerance with $L = 10$, $K =4$, $D=1$ km, and $p_{\max} = 200$ mW.}
\label{fig6}
\end{figure}

\textcolor{black}{Fig. \ref{fig6} shows the uplink training time against the lag tolerance with lag percent $\nu = 85$. As expected, the uplink training time first decreases along with the increase of the number of blocks, which is resulted from that data exchange is less often. Then, the uplink training time increases as the SE of UEs are declined. Besides, the efficiency of our power control method is obvious. When the number of blocks is 8, applying the proposed power control method leads to a 42\% reduction in terms of uplink training time.}

\begin{figure}[t!]
\centering
\includegraphics[scale=0.6]{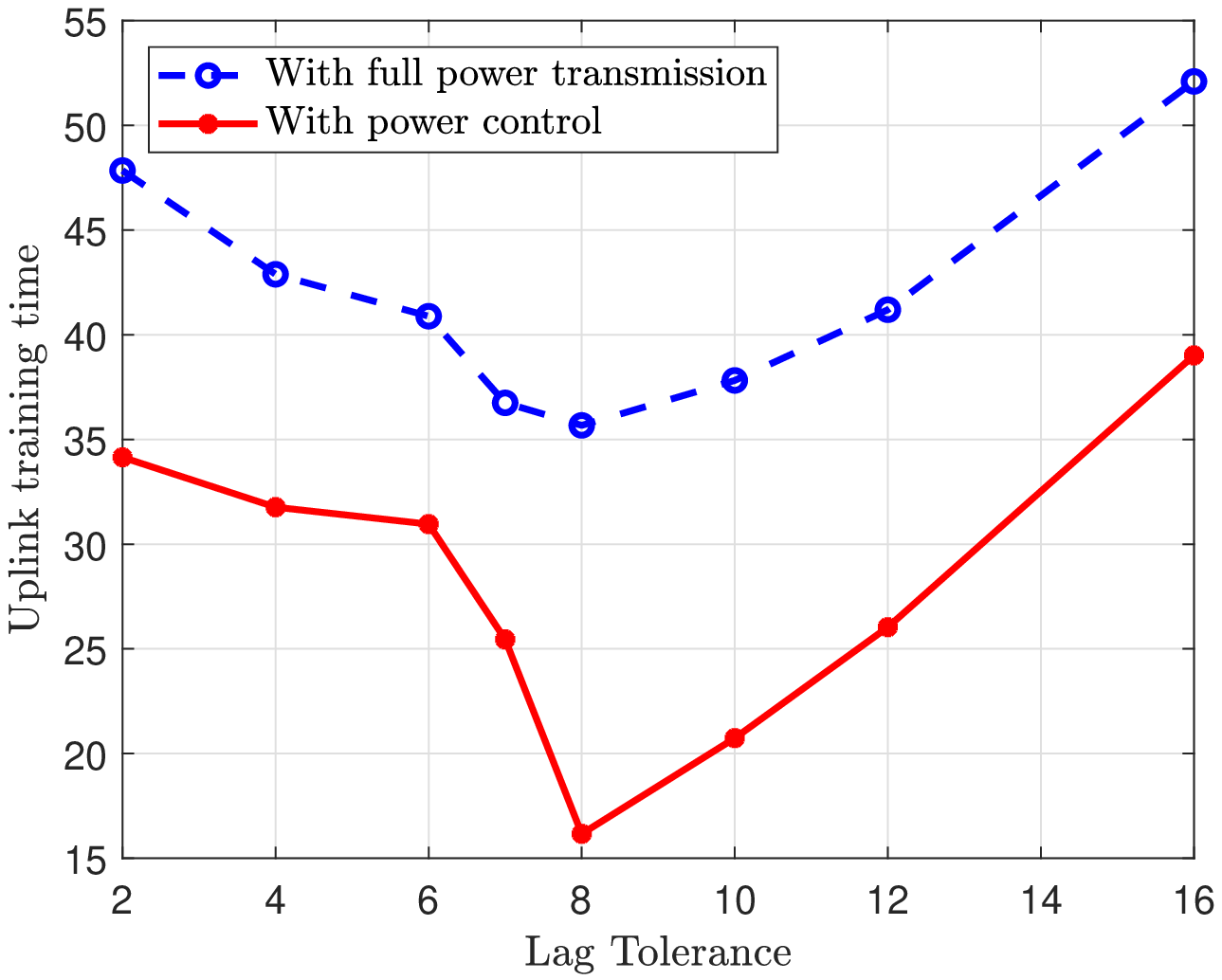}
\caption{Uplink training time against $v$\% with $L = 10$, $K =4$, $D=1$ km, and $p_{\max} = 200$ mW.}
\label{fig7}
\end{figure}

\textcolor{black}{Fig. \ref{fig7} shows the uplink training time against the lag percent with a lag tolerant is equal to 4, which reflects the number of UEs served by the APs in each round of communication. It can be observed from Fig. \ref{fig7} that when $\nu < 80\%$, the uplink training time decreases as $\nu$ increases. The reason is that the straggler UEs who significantly slow down the whole FL process is not served by the APs in every communication iteration. However, when $\nu$ continues to reduce, the uplink training time begins to increase, since the number of users served in each round of communication decreases, which leads to the decreases of SE that in turn increases the training time. Therefore, Fig. \ref{fig7} shows that choosing an appropriate value of $\nu$ can effectively reduce the training time.}

\section{Conclusions}
In this paper, we studied the DP in wireless FL enabled by CF mMIMO systems with low-resolution ADCs and DACs. By introducing the quantization noise as the DP mechanism, we derived an expression of the privacy preservation condition and provided convergence analysis for the proposed model. Targeting at the uplink training time minimization, we jointly optimized the transmit power and data rate under different DP protection levels. The simulation results showed that our proposed power control method can effectively reduce the uplink training time in all considered cases. In order to mitigate the effect of straggler, we proposed an asynchronous FL protocol which incorporates a UE selection algorithm based on the large-scale fading coefficient decoupling the CPU and the selected UEs for a reduction of uplink training time and for tackling the tradeoff between faster convergence and lower communication overhead. \textcolor{black}{To further improve the system performance, some future extensions can be considered, e.g., channel allocation.}

\bibliographystyle{IEEEtran}
\bibliography{IEEEabrv,Ref_FL}

\end{document}